\documentstyle[eqsecnum,aps,epsf]{revtex}

\begin{document}
\draft
\twocolumn
\title{Coherent x-ray scattering from manganite charge and orbital domains\\}
\author{C.S. Nelson,$^1$\cite{byline} J.P. Hill,$^1$ Doon Gibbs,$^1$ F. Yakhou,$^2$ F. Livet,$^3$ Y. Tomioka,$^4$ T. Kimura,$^4$ and Y. Tokura$^{4,5}$\\}
\address{$^1$Department of Physics, Brookhaven National Laboratory, Upton, NY  11973-5000\\$^2$European Synchrotron Radiation Facility, B.P. 220-38043, Grenoble Cedex, France \\$^3$LTPCM-ENSEEG-INPG, UMR-CNRS No. 5614, B.P. 75-38402, St. Martin d'H$\grave{e}$res Cedex, France \\$^4$ Correlated Electron Research Center (CERC), AIST, Tsukuba 305-0033, Japan\\$^5$Department of Applied Physics, University of Tokyo, Tokyo 113-8656, Japan\\}
\date{\today}
\maketitle
\begin{abstract}
We report coherent x-ray scattering studies of charge and orbital domains in manganite systems.  The experiments were carried out on LaMnO$_3$ and Pr$_{0.6}$Ca$_{0.4}$MnO$_3$, with the incident photon energy tuned near the Mn K edge.  At room temperature, the orbital speckle pattern of LaMnO$_3$ was observed to be constant over a timescale of at least minutes, which is indicative of static orbital domains on this timescale.  For Pr$_{0.6}$Ca$_{0.4}$MnO$_3$, both charge and orbital speckle patterns were observed.  The observation of the latter rules out the presence of fast orbital fluctuations, while long time series data--- on the order of several minutes--- were suggestive of slow dynamic behavior.  In contrast, the charge order speckle patterns were static.
\end{abstract}

\narrowtext

Coherent x-ray scattering is a powerful tool for the study of domain structure in condensed matter systems because of its sensitivity to the exact spatial arrangement of the domains.  That is, upon illumination by a coherent beam, a sample composed of a mosaic of domains introduces a set of different phases into the scattered beam.  The interference of these phases, which are related to the positions of the individual domains, gives rise to a ``speckle'' pattern.  In principle, this speckle pattern can be inverted in order to obtain a real-space image of the domains.  Such a reconstruction is quite challenging, but has been accomplished in systems consisting of a small number of domains \cite{robinson1,robinson2}.  

Another promising use of coherent x-ray scattering is in the study of dynamics.  If the spatial arrangement of the domains changes with time, the phases of the scattered beam will be affected and the speckle pattern will also change.  Measurements of these changes can therefore be used to study the dynamics of domain fluctuations.  In the x-ray regime, such measurements have been used to study order-disorder transitions \cite{brauer}, the dynamics of colloids \cite{dierker,thurn,riese} and block copolymer micelles \cite{mochrie}, liquid crystal fluctuations \cite{fera}, alloy decomposition kinetics \cite{livet}, and magnetic domain fluctuations \cite{yakhou}.

The characteristics of coherent x-ray scattering, as described above, are well-matched to an outstanding problem within the field of manganite physics--- specifically, the problem of the orbital domain state exhibited by many perovskite manganites of the general formula R$_{1-x}$M$_x$MnO$_3$, where R and M are rare-earth and alkaline cations, respectively, at or near half-doping (i.e., x $\approx$ 0.5).  At low temperatures, this class of materials is believed to order into a structure involving arrangements of the 3d spin and electronic charge, together with a well-defined 3d orbital occupation.  In this so-called CE-type phase, which is shown in Figure 1, the ordering consists of ferromagnetic zigzags, coupled antiferromagnetically; a checkerboard charge arrangement of Mn$^{3+}$ and Mn$^{4+}$ ions, stacked in-phase along the c-axis; and orbital ordering with a concomitant lattice distortion that doubles the size of the unit cell along one direction.  While this complicated ordering was determined via neutron scattering measurements nearly 50 years ago \cite{wollan}, a theoretical grounding for the CE-type phase is still being developed \cite{solovyev,brink,yunoki,gu,popovic}.
	
Interest in the CE-type phase has been rekindled in part because of its sensitivity to perturbations.  For example in some materials exhibiting this phase, application of a magnetic field destroys the charge, orbital, and antiferromagnetic ordering and drives the material from an insulating into a ferromagnetic metallic state--- the phenomenon known as colossal magnetoresistance \cite{tokura}.  Recently, attention has focused on the domain states of the various kinds of order in the CE-type phase.  One particularly intriguing result of these studies is that the orbital and Mn$^{3+}$ magnetic sublattices have been observed to exhibit only short-range order--- each forming a domain state with a correlation length on the order of $\sim$100 $\AA$.  In contrast, the charge and magnetic order of the Mn$^{4+}$ sublattice achieve long-range order \cite{radaelli,zim_long,valery}.  Reconciling these results, it has been shown \cite{radaelli,zim_long} that orbital domain walls will destroy the coherence of the magnetic structure on the Mn$^{3+}$ sites only, leaving the Mn$^{4+}$ magnetic sublattice intact, if the checkerboard charge arrangement is also retained.  Thus, the central question is what is the origin of these orbital domain walls, and more specifically, what determines the length scale of the orbital domains?

One possibility, first suggested by Millis \cite{millis}, lies in the misfit strain that arises from the lattice distortion associated with orbital ordering and grows with the size of the domains.  Recent calculations pertaining to the accommodation of such strain in another 3d transition metal system--- the cuprates--- result in a characteristic length scale of stripe domains on the order of a hundred lattice spacings \cite{billinge}, which is of the same order of magnitude as the orbital correlation length observed in the CE-type phase of the perovskite manganites.  Similar arguments have also been made by Khomskii and Kugel \cite{kandk}.  A second possibility is that the orbital domain walls are pinned by fluctuations in the rare earth and alkaline cation distribution.  Assuming that the cation distribution is quenched, such a mechanism might be expected to lead to the formation of orbital domain walls at the same location upon thermal cycling.

Another, related question concerns the time dependence of the orbital domains--- i.e., are they static or dynamic?  The short-range order exhibited by the orbital domain state is reminiscent of a glass-like phase.  It is possible that the formation of this orbital domain state represents a freezing of the dynamics in a glass-like transition--- perhaps near the charge and orbital order transition temperature.  Indeed, such a transition has recently been observed in the dynamics of correlated polarons in the insulating phase of a bilayer manganite with a low temperature, ferromagnetic metallic state \cite{argyriou}.  Indirect support of such a picture in the present system follows from the observation of slow dynamics--- on the timescale of a few seconds--- near an electron beam-induced charge ordering transition in La$_{0.225}$Pr$_{0.4}$Ca$_{0.375}$MnO$_3$ \cite{podzorov}, while neutron scattering measurements of static short-range ordered magnetic correlations imply that the orbital domains are static on the timescale of picoseconds below the N\'{e}el temperature \cite{kajimoto}.  However to date, there is very little direct information on the dynamics of orbital domains at larger timescale and higher temperatures.

In order to address the questions posed above and thereby shed light on the intriguing orbital domain state exhibited by the CE-type phase of perovskite manganites, we have carried out a coherent x-ray scattering study of a well-characterized single-crystal of Pr$_{0.6}$Ca$_{0.4}$MnO$_3$ \cite{zim_long}.  We have observed coherent x-ray scattering from both charge and orbital domains.  In addition, we have investigated the orbital domains of a single-crystal of LaMnO$_3$.  Both the orbital domains in LaMnO$_3$ and the charge domains in Pr$_{0.6}$Ca$_{0.4}$MnO$_3$ were observed to be static over a timescale of at least minutes.  Studies of the orbital speckle pattern of the latter material--- while ruling out relatively fast dynamics--- were quantitatively inconclusive and will require additional experiments.

The coherent x-ray scattering measurements were carried out on the ID20 beamline at the European Synchrotron Radiation Facility (ESRF).  This beamline is equipped with a Si(111) double-crystal monochromator, which was used to set the incident photon energy to 6.55 keV.  This energy was chosen to take advantage of the resonant enhancement in the scattering observed at the charge and orbital wavevectors upon tuning to the vicinity of the Mn K edge (see inset to Figure 2) \cite{murakami1,murakami2}.  The undulator source size of 0.8$\times$0.3 mm$^{2}$ (H$\times$V) FWHM resulted in transverse coherence lengths of $\sim$10$\times$30 $\mu$m$^{2}$ (H$\times$V) at a distance of 45 m from the source.  An isotropic coherent beam was selected using slits set to a gap of 60$\times$60 $\mu$m$^{2}$ and a 10 $\mu$m diameter pinhole.  The incident, longitudinal coherence length, which is determined by the 1.4$\times$10$^{-4}$ resolution of the monochromator at the Mn K edge, was $\sim$1 $\mu$m.  This value is greater than the maximum path length difference in the experiment--- even at the largest scattering angle of $\sim$30$^{\circ}$--- due to the reduced absorption length at photon energies near the absorption edge.  The beam therefore satisfies the requirements for observing coherent diffraction.  The coherent beam intensity incident on the sample was 1.4$\times$10$^{9}$ photons/s. 

Initial measurements were conducted on a single-crystal of LaMnO$_{3}$ that was grown using floating zone techniques at CERC.  This material exhibits C-type orbital order below a temperature of $\sim$800 K \cite{murakami2}.  The sample was mounted in air in a horizontally-scattering diffractometer.  A NaI scintillation detector with a 2 mm diameter active area was used for alignment purposes, and the speckle patterns were subsequently collected using a CCD camera (with 384$\times$576, 22$\times$22 $\mu$m$^{2}$ pixels), at a position 1.8 m from the sample.  At the (010) orbital order peak, the scattering intensity was $\sim$800 counts/s using the scintillation detector.

Examples of three-pixel-wide slices through two images measured at the orbital order wavevector, and their residual, are displayed in the upper and lower parts of Figure 2, respectively.  In both slices, the contrast between the peaks and valleys is dramatic.  This contrast is not a result of limited statistics, but is indicative of the interference of coherent scattering from individual orbital domains.  The widths of the peaks--- or speckles--- is also consistent with the 10 $\mu$m diameter of the pinhole placed in front of the sample.  That is, the speckles produced by coherent x-ray scattering should be the size of the Fraunhofer diffraction pattern of the coherent beam-defining optic.  In our set-up, this corresponds to $\sim$2 pixels, as observed.  Note that this is also the justification for binning in the orthogonal direction.  

Most significantly, the two slices shown in Figure 2 are nearly identical, as illustrated by the near-zero value of the residual.  This result demonstrates that the interference pattern--- and therefore the orbital domains in LaMnO$_3$--- are essentially static over a timescale of at least minutes.  The data shown in Figure 2 also provide an estimate of the average size of the orbital domains.  We note in passing that the resolution in this experiment, which is determined by the pixel size of the CCD and the sample-to-detector distance, is better than the previous high-resolution x-ray scattering measurements conducted on this material that reported resolution-limited peaks \cite{zim_to}.  Here we find the orbital domain size, which is inversely proportional to the reciprocal of the envelope width of the entire speckle pattern, to be 4000--7000 $\AA$.

After the successful demonstration of static orbital speckle scattering in LaMnO$_3$, a single-crystal of Pr$_{0.6}$Ca$_{0.4}$MnO$_{3}$, which exhibits CE-type charge and orbital order below a temperature of $\sim$240 K, was investigated.  The x = 0.4 doping was chosen because its orbital correlation length of 320 $\pm$ 10 $\AA$ is the longest observed in the CE-type phase of this system \cite{zim_long,hill}.  As a result, the structure in the speckle patterns is minimized, since the number of speckles is proportional to the number of orbital domains illuminated by the incident beam.  The sample was mounted in a $^{4}$He cryostat and cooled to a temperature of 150 K.  The NaI scintillation detector was again used to align the sample, and images were then collected with the CCD camera.  Examples of images at the (0 2.5 0) orbital order and (030) charge order peaks are displayed in Figure 3.  The orbital order image is observed to be much larger than the charge order images as a result of the shorter orbital correlation length.  With regard to the charge order, we note the near indistinguishable appearance of the charge speckle patterns in (b) and (c).  These two images were collected successively, each over a duration of 100 s, and their appearance indicates that the charge domains are static at this temperature, over a timescale of at least minutes.

A series of speckle patterns were next collected in an attempt to study the temperature dependence of the charge and orbital domains.  Speckle patterns were collected between the temperatures of 150 and 230 K, above which the scattering intensity from the orbital order was too weak to observe over reasonable collection times.  Examples of these temperature dependence measurements for the charge and orbital speckle patterns are displayed in Figure 4(a,b) and (c,d), respectively, in each case at temperatures of 150 and 200 K.

Focusing first on the charge order images in Figure 4(a,b), we note that the dramatic change in the structure of the slices between the two temperatures is due to the fact that the sample moves slightly in the beam as the temperature changes.  As a result, different grains within the sample are illuminated, producing a different speckle pattern.  However, at a single temperature, successive slices through the charge speckle patterns are found to be remarkably similar in appearance.  The data shown for a temperature of 150 K (Figure 4(a)) are slices taken through the images displayed in Figure 3(b) and 3(c).  As remarked earlier, the two images look very similar, and this is reflected in the close correspondence of the slices.  Further, images collected within minutes of each other at a temperature of 200 K (Figure 4(b)) are also observed to be quite similiar.  Taken together, these results are indicative of static charge domains in Pr$_{0.6}$Ca$_{0.4}$MnO$_3$ over a timescale of at least minutes, in the temperature range of 150--200 K.

Moving on to the orbital speckle scattering, which is displayed in Figure 4(c,d), the results are less conclusive.  This is due to the weaker orbital scattering intensity--- compared to that of the charge order--- and the shorter correlation length that results in an expanded image.  To partially counteract these difficulties, longer collection times were utilized.  Slices through successive images collected over 750 s at temperatures of 150 and 200 K are shown in Figure 4(c) and 4(d), respectively.  At neither temperature is there a close correspondence between successive images, suggesting that the orbital domains may be dynamic.  However, the presence of fine structure in all four slices indicates that any dynamic behavior is relatively slow--- on a timescale of several minutes.  This follows from the fact that if the orbital domains had correlation times much shorter than minutes, then the structure of a given slice would be washed out into a smooth distribution.  Therefore in this temperature range, we can rule out the presence of orbital fluctuations on the timescale of picoseconds, as inferred by Kajimoto {\it et al.} \cite{kajimoto}.  In terms of the temperature dependence of any orbital domain dynamics, though, quantitative results are difficult to obtain from the current data set due to a lack of statistical significance.

In Pr$_{0.6}$Ca$_{0.4}$MnO$_3$, we observe static charge domains over a timescale of at least minutes, while the speckle scattering from the orbital domains is suggestive of dynamic behavior.  This latter result leads us to speculate that the orbital domain state is, in fact, glass-like \cite{hill}.  This is consistent with the lack of long-range order, and in addition, suggests a framework for investigations based on the wealth of experimental and theoretical work on other glass-like systems.  An obvious analog to this hypothetical, orbital glass-like state in the CE-type phase is the class of materials known as spin glasses \cite{binder}.  In spin glasses, static structural disorder or magnetic frustration precludes the development of long-range magnetic order (though the typical correlation lengths are an order of magnitude smaller than those observed in the hypothetical, orbital glass-like phase).  Instead, below a freezing temperature, the magnetic moments are oriented in arbitrary and fixed--- over macroscopic timescales--- directions.  Theoretical models of spin glasses require frustration, in that no spin configuration simultaneously minimizes competing terms in the Hamiltonian.  For the case of orbital order discussed here, such frustration could arise from competition between the cooperative Jahn-Teller distortion that constitutes the orbital ordering and fluctuations in the local Jahn-Teller distortion due to quenched disorder of the cation distribution.  That is, the fluctuations could interact with the cooperative Jahn-Teller distortion and limit the correlation length of the orbital ordering.  

Using the analogy between the hypothetical orbital glass-like state in the CE-type phase and spin glasses, directions for future studies become clear.  First, it must be determined whether or not the orbital domains exhibit a true glass transition between solid-like and liquid-like phases.  Evidence of this could be obtained via further coherent x-ray scattering studies.  A particularly promising approach might be to tune the incident photon energy to the Mn L$_2$ and L$_3$ edges and use soft, coherent x-ray scattering where much larger resonant enhancements are expected \cite{castleton} and significant coherent fluxes can be obtained (e.g. 10$^{12}$ photons/s \cite{hu}).  Using coherent x-ray scattering, the observation of a dynamic transition in the orbital speckle scattering at some freezing temperature would provide strong support for a glass-like picture.  Following this, relaxation times of the orbital domain state should be characterized.  In spin glasses, the relaxation has a very broad spectrum ranging from macroscopic values below the freezing temperature to tens of picoseconds at temperatures well above this transition.  In the studies of the orbital domain state of the CE-type phase reported here, only possibly macroscopic timescales for relaxation have been observed.  In the future, other routes for the characterization of relaxation times might include studies of properties that couple to the orbital order parameter--- for example, by taking advantage of the connection between the lattice distortion and orbital order.

In conclusion, we have used coherent x-ray scattering to study orbital domains in LaMnO$_3$ and Pr$_{0.6}$Ca$_{0.4}$MnO$_3$, and charge domains in the latter material.  In LaMnO$_3$, the orbital domains are static at room temperature over a timescale of at least minutes.  In Pr$_{0.6}$Ca$_{0.4}$MnO$_3$, the charge domains are static over the same macroscopic timescale, while the behavior of the orbital domains is suggestive of dynamic behavior on a relatively long timescale that may be consistent with the formation of a glass-like state.  This suggestion of glass-like behavior is extremely intriguing and warrants future study.

\vspace{0.125in}

The work at Brookhaven was supported by the U.S. Department of Energy, Division of Materials Science, under Contract No. DE-AC02-98CH10886.  We thank Carsten Detlefs for help during the measurements.

\begin{figure}
\vspace{0.25in}
\epsfxsize=.3\textwidth
\centerline{\epsffile{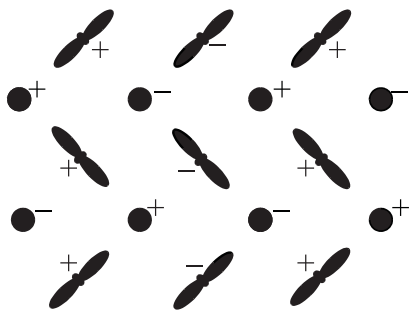}}
\vspace{0.25in}
\caption{Schematic diagram of the CE-type structure, in the $a-b$ plane.  Closed circles represent Mn$^{4+}$ ions, elongated figure-eights represent the occupied e$_g$ (3d$_{z^2-r^2}$-type) orbital of Mn$^{3+}$ ions, and $+$ and $-$ indicate the magnetic ordering.}
\end{figure} 

\begin{figure}
\vspace{0.25in}
\epsfxsize=.5\textwidth
\centerline{\epsffile{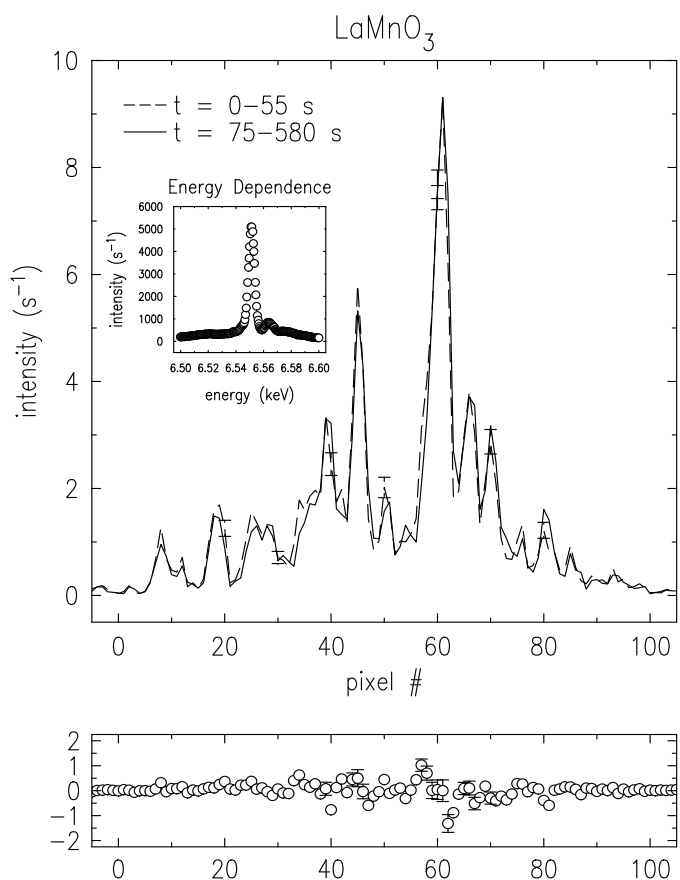}}
\vspace{0.25in}
\caption{Linear slices through the (010) orbital peak in LaMnO$_3$ (upper).  The two slices were taken successively and are 3 pixels wide in the perpendicular direction (the approximate speckle size).  Data were collected at room temperature, and for clarity, error bars are only displayed at every tenth pixel.  The difference between the two slices is shown in the lower panel.  Inset shows the energy dependence of the scattered intensity at the orbital order wavevector.}
\end{figure}

\begin{figure}
\epsfxsize=.5\textwidth
\centerline{\epsffile{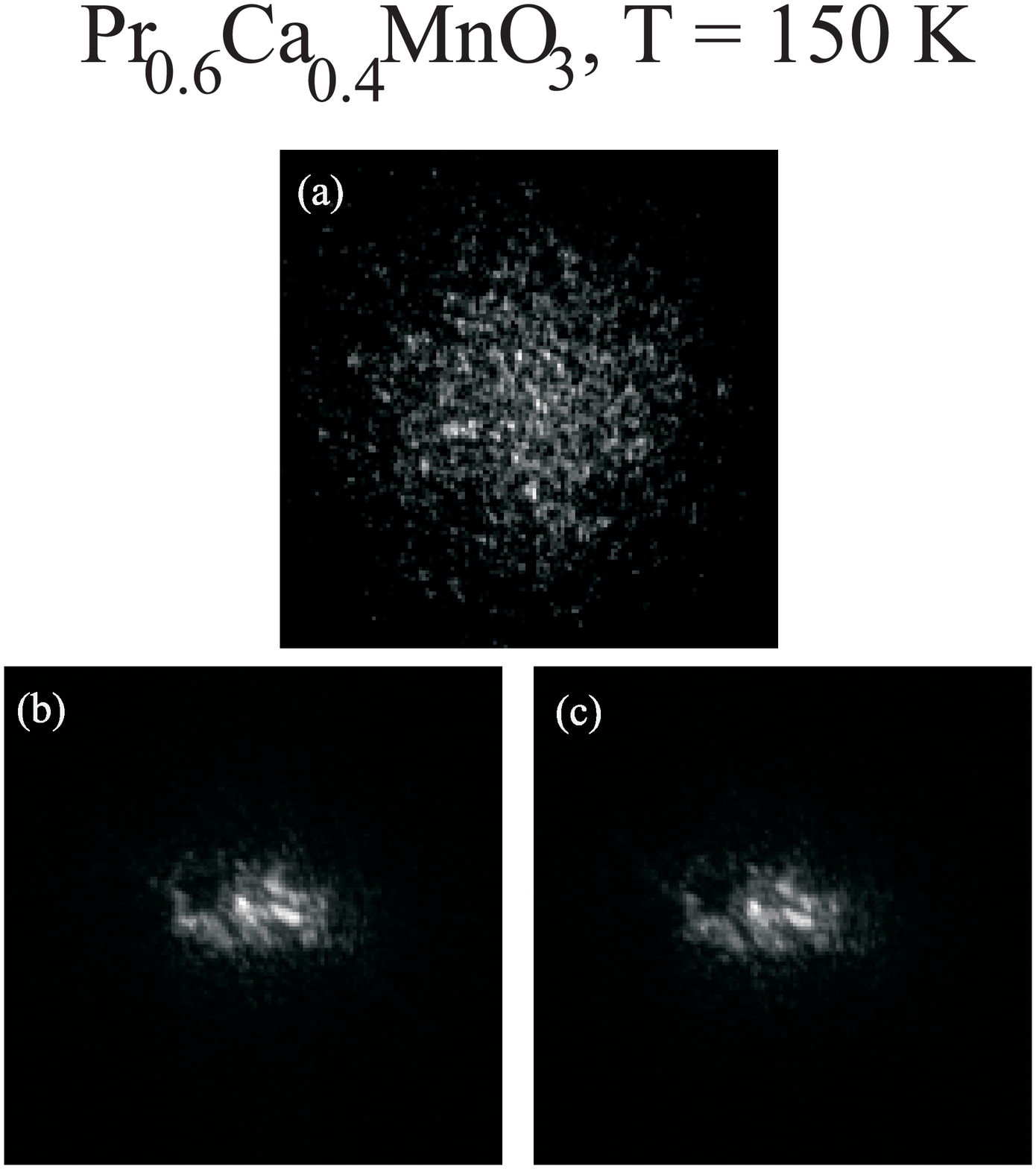}}
\vspace{0.25in}
\caption{Images of orbital (a) and charge (b, c) order speckle scattering measured at a temperature of 150 K.  The image in (a) consists of 101 frames, measured for 15 s per frame; the images in (b) and (c) each consist of 101 frames, measured for 1 s per frame.  Note that the two images in (b) and (c) were collected within minutes of each other.}
\end{figure}

\begin{figure}
\epsfxsize=.5\textwidth
\centerline{\epsffile{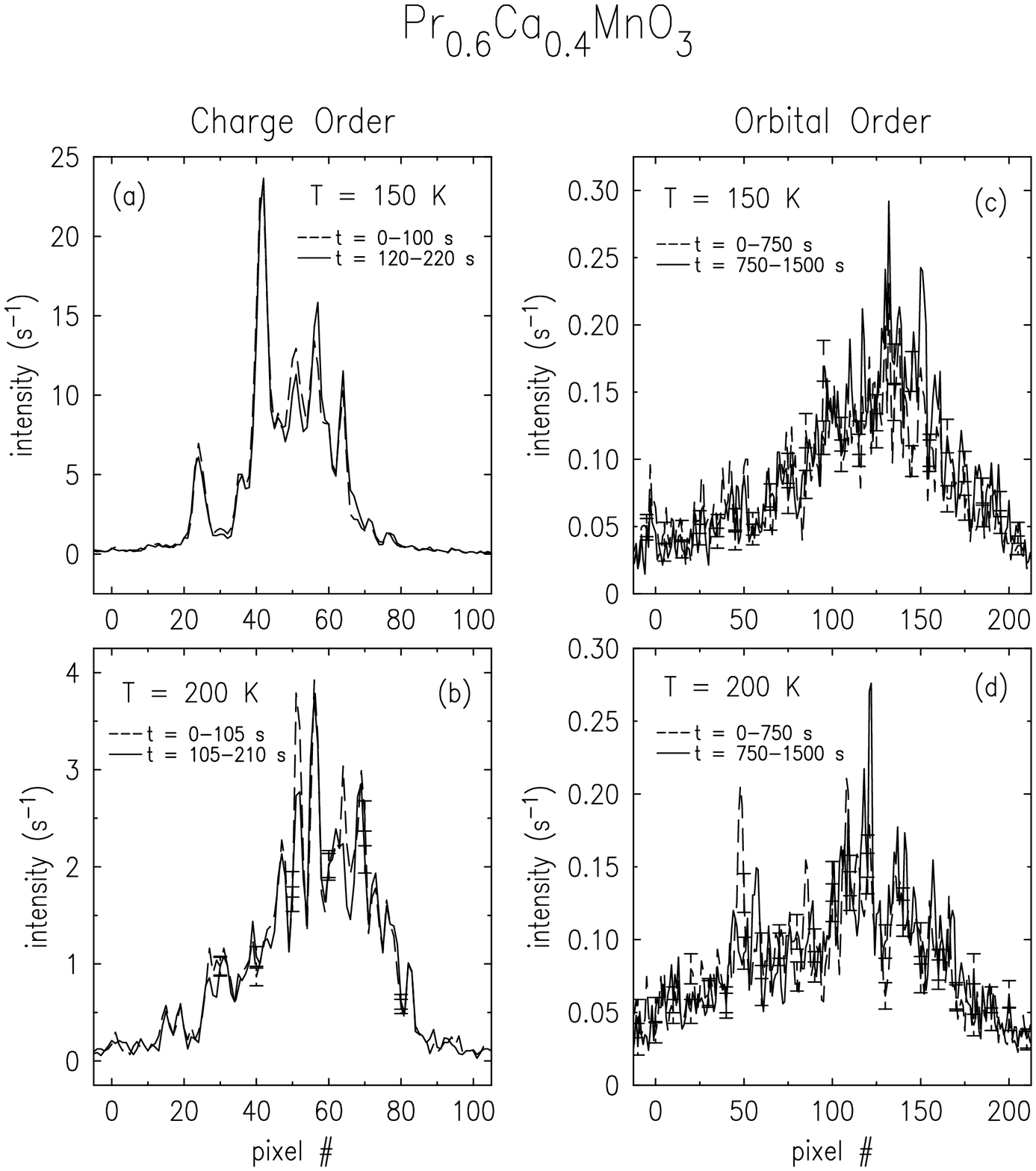}}
\vspace{0.25in}
\caption{Linear slices through images of charge/orbital speckle scattering from Pr$_{0.6}$Ca$_{0.4}$MnO$_{3}$, measured at 150 K (a/c) and 200 K (b/d).  Slices are three pixels wide in the direction perpendicular to the scattering plane.  For clarity, error bars are only displayed at every tenth pixel.}
\end{figure}

\end{document}